\documentclass[runningheads]{llncs}
\usepackage[T1]{fontenc}
\usepackage{graphicx}
\usepackage{todonotes}
\usepackage{cite}
\usepackage{amsmath,amssymb,amsfonts}
\usepackage{textcomp}
\usepackage{wrapfig}
\usepackage{algorithmic}
\usepackage{graphicx}
\usepackage{textcomp}
\usepackage{xcolor, cancel}
\usepackage{array}
\usepackage{tabularx}
\usepackage{multirow, multicol}
\usepackage{threeparttable}
\usepackage{color, colortbl}
\usepackage{caption,subcaption}
\usepackage{balance}
\usepackage{hyperref}
\usepackage{cleveref}
\usepackage{amsmath}
\usepackage{enumitem}
\usepackage[normalem]{ulem}
\usepackage{lipsum}
\usepackage{float}
\usepackage{soul}

\crefformat{section}{\S#2#1#3}

\definecolor{Gray}{gray}{0.90}

\makeatletter
\renewcommand{\fnum@figure}{Figure. \thefigure}
\makeatother

\newcolumntype{C}{>{\centering\arraybackslash}X}

\begin{document}
\title{Analyzing Resource Utilization in an HPC System: A Case Study of NERSC's Perlmutter}
\titlerunning{A Case Study of NERSC's Perlmutter}
%

%

\author{Jie Li\inst{1}\orcidID{0000-0002-5311-3012}\and
George Michelogiannakis\inst{2}\orcidID{0000-0003-3743-6054}\and
Brandon Cook\inst{2}\orcidID{0000-0002-4203-4079}\and
Dulanya Cooray\inst{3}\orcidID{0009-0000-1727-6298}\and
Yong Chen\inst{1}\orcidID{0000-0002-9961-9051}}
\authorrunning{J. Li et al.}

%
\institute{Texas Tech University, Lubbock, TX 79409, USA \\
\email{\{jie.li, yong.chen\}@ttu.edu}\and
Berkeley Lab, Berkeley, CA 94720, USA \\
\email{\{mihelog, bgcook\}@lbl.gov}\and
University of California, Berkeley, CA 94720, USA\\
\email{\{dulanya\}@berkeley.edu}}

\maketitle              
%


\begin{abstract}
Resource demands of HPC applications vary significantly. However, it is common for HPC systems to primarily assign resources on a per-node basis to prevent interference from co-located workloads. This gap between the coarse-grained resource allocation and the varying resource demands can lead to HPC resources being not fully utilized. In this study, we analyze the resource usage and application behavior of NERSC's Perlmutter, a state-of-the-art open-science HPC system with both CPU-only and GPU-accelerated nodes. Our one-month usage analysis reveals that CPUs are commonly not fully utilized, especially for GPU-enabled jobs. Also, around 64\% of both CPU and GPU-enabled jobs used 50\% or less of the available host memory capacity. Additionally, about 50\% of GPU-enabled jobs used up to 25\% of the GPU memory, and the memory capacity was not fully utilized in some ways for all jobs. While our study comes early in Perlmutter's lifetime thus policies and application workload may change, it provides valuable insights on performance characterization, application behavior, and motivates systems with more fine-grain resource allocation.


\end{abstract}

\section{Introduction}
\label{section:introduction}

In the past decade, High-Performance Computing (HPC) systems shifted from traditional clusters of CPU-only nodes to clusters of more heterogeneous nodes, where accelerators such as GPUs, FPGAs, and 3D-stacked memories have been introduced to increase compute capability~\cite{kindratenko2011trends}. Meanwhile, the collection of open-science HPC workloads is particularly diverse and recently increased its focus on machine learning and deep learning~\cite{gil2014amplify}. Heterogeneous hardware combined with diverse workloads that have a wide range of resource requirements makes it difficult to achieve efficient resource management. Inefficient resource management threatens to not fully utilize expensive resources that can rapidly increase capital and operating costs. Previous studies have shown that the resources of HPC systems are often not fully utilized, especially memory~\cite{panwar2019quantifying, peng2020memory, michelogiannakis2022case}. 

NERSC's Perlmutter also adopts a heterogeneous design to bolster performance, where CPU-only nodes and GPU-accelerated nodes together provide a three to four times performance improvement over Cori~\cite{perlmutter, cori}, making Perlmutter rank 8th in the Top500 list as of December 2022. However, Perlmutter serves a diverse set of workloads from fusion energy, material science, climate research, physics, computer science, and many other science domains~\cite{nersc_workload}. In addition, it is useful to gain insight into how well users are adapting to Perlmutter's heterogeneous architecture.

Consequently, it is desirable to understand how system resources in Perlmutter are used today. The results of such an analysis can help us evaluate current system configurations and policies, provide feedback to users and programmers, offer recommendations for future systems, and motivate research in new architectures and systems.
In this work, we focus on understanding CPU utilization, GPU utilization, and memory capacity utilization (including CPU host memory and GPU memory) on Perlmutter. These resources are expensive, consume significant power, and largely dictate application performance.

In summary, our contributions are as follows:

\begin{itemize}
  \item We conduct a thorough utilization study of CPUs, GPUs, and memory capacity in Perlmutter, a top 8 state-of-the-art HPC system that contains both CPU-only and GPU-accelerated nodes. We discover that both CPU-only and GPU-enabled jobs usually do not fully utilize key resources.
  \item We find that host memory capacity is largely not fully utilized for memory-balanced jobs, while memory-imbalanced jobs have significant temporal and/or spatial memory requirements.
  \item We show a positive correlation between job node hours, maximum memory usage, as well as temporal and spatial factors.
  \item Our findings motivate future research such as resource disaggregation, job scheduling that allows job co-allocation, and research that mitigates potential drawbacks from co-locating jobs.
\end{itemize}

\section{Related Work}
\label{section:related work}
Many previous works have utilized job logs and correlated them with system logs to analyze job behavior in HPC systems~\cite{oliner2007supercomputers, zheng2011co, di2017logaider, madireddy2017analysis, gupta2017failures}.
For example, Zheng et al. correlated the Reliability, Availability, and Serviceability (RAS) logs with job logs to identify job failure and interruption characteristics~\cite{zheng2011co}. Other works utilize performance monitoring infrastructure to characterize application and system performance in HPC~\cite{ji2017understanding, turner2018survey, patel2019revisiting, wang2019learning, li2020monster, peng2021holistic, michelogiannakis2022case}. In particular, the paper presented by Ji et al. analyzed various application memory usage in terms of object access patterns~\cite{ji2017understanding}. Patel et al. collected storage system data and performed a correlative analysis of the I/O behavior of large-scale applications~\cite{patel2019revisiting}. The resource utilization analysis of the Titan system~\cite{wang2019learning} summarized the CPU and GPU time, memory, and I/O utilization across a five-year period. Peng et al. focused on the memory subsystem and studied the temporal and spatial memory usage in two production HPC systems at LLNL~\cite{peng2021holistic}. Michelogiannakis et al.~\cite{michelogiannakis2022case} performed a detailed analysis of key metrics sampled in NERSC's Cori to quantify the potential of resource disaggregation in HPC. 

System analysis provides insights into resource utilization and therefore drives research on predicting and improving system performance~\cite{xie2017predicting, das2018desh, panwar2019quantifying, peng2020memory}. Xie et.al developed a predictive model for file system performance on the Titan supercomputer~\cite{xie2017predicting}. Desh~\cite{das2018desh}, proposed by Das et al., is a framework that builds a deep learning model based on system logs to predict node failures. Panwar et al. performed a large-scale study of system-level memory utilization in HPC and proposed exploiting unused memory via novel architecture support for OS~\cite{panwar2019quantifying}. Peng et al. performed a memory utilization analysis of HPC clusters and explored using disaggregated memory to support memory-intensive applications~\cite{peng2020memory}.

\section{Background}
\label{section:background}

\subsection{System Overview}
\label{section:perlmutter}
NERSC's latest system, Perlmutter~\cite{perlmutter}, contains both CPU-only nodes and GPU-accelerated nodes with CPUs. Perlmutter has 1,536 GPU-accelerated nodes (12 racks, 128 GPU nodes per rack) and 3,072 CPU-only nodes (12 racks, 256 CPU nodes per rack). These nodes are connected through HPE/Cray's Slingshot Ethernet-based high performance network. Each GPU-accelerated node features four NVIDIA A100 Tensor Core GPUs and one AMD ``Milan'' CPU. The memory subsystem in each GPU node includes 40 GB of HBM2 per GPU and 256 GB of host DRAM. Each CPU-only node features two AMD ``Milan'' CPUs with 512 GB of memory. Perlmutter currently uses SLURM version 21.08.8 for resource management and job scheduling. Most users submit jobs to the regular queue that has no maximum number of nodes and a maximum allowable duration of 12 hours.

The workload served by the NERSC systems includes applications from a diverse range of science domains, such as fusion energy, material science, climate research, physics, computer science, and more~\cite{nersc_workload}. From the over 45-year history of the NERSC HPC facility and 12 generations of systems with diverse architectures, the traditional HPC workloads evolved very slowly despite the substantial underlying system architecture evolution~\cite{michelogiannakis2022case}. However, the number of deep learning and machine learning workloads across different science disciplines has grown significantly in the past few years~\cite{thomas2021monitoring}. Furthermore, in our sampling time, Perlmutter was operating in parallel with Cori. Thus, the NERSC workload was divided among the two machines and Perlmutter's workload may change once Cori retires. Therefore, while our study is useful to (i) find the gap between resource provider and resource user and (ii) extract insights early in Perlmutter's lifetime to guide future policies and procurement, as in any HPC system the workload may change in the future. Still, our methodology can be reused in the future and on different systems.

\subsection{Data Collection}
\label{section:data collection}

\begin{figure}[t]
    \centering
    \includegraphics[width=0.70\linewidth]{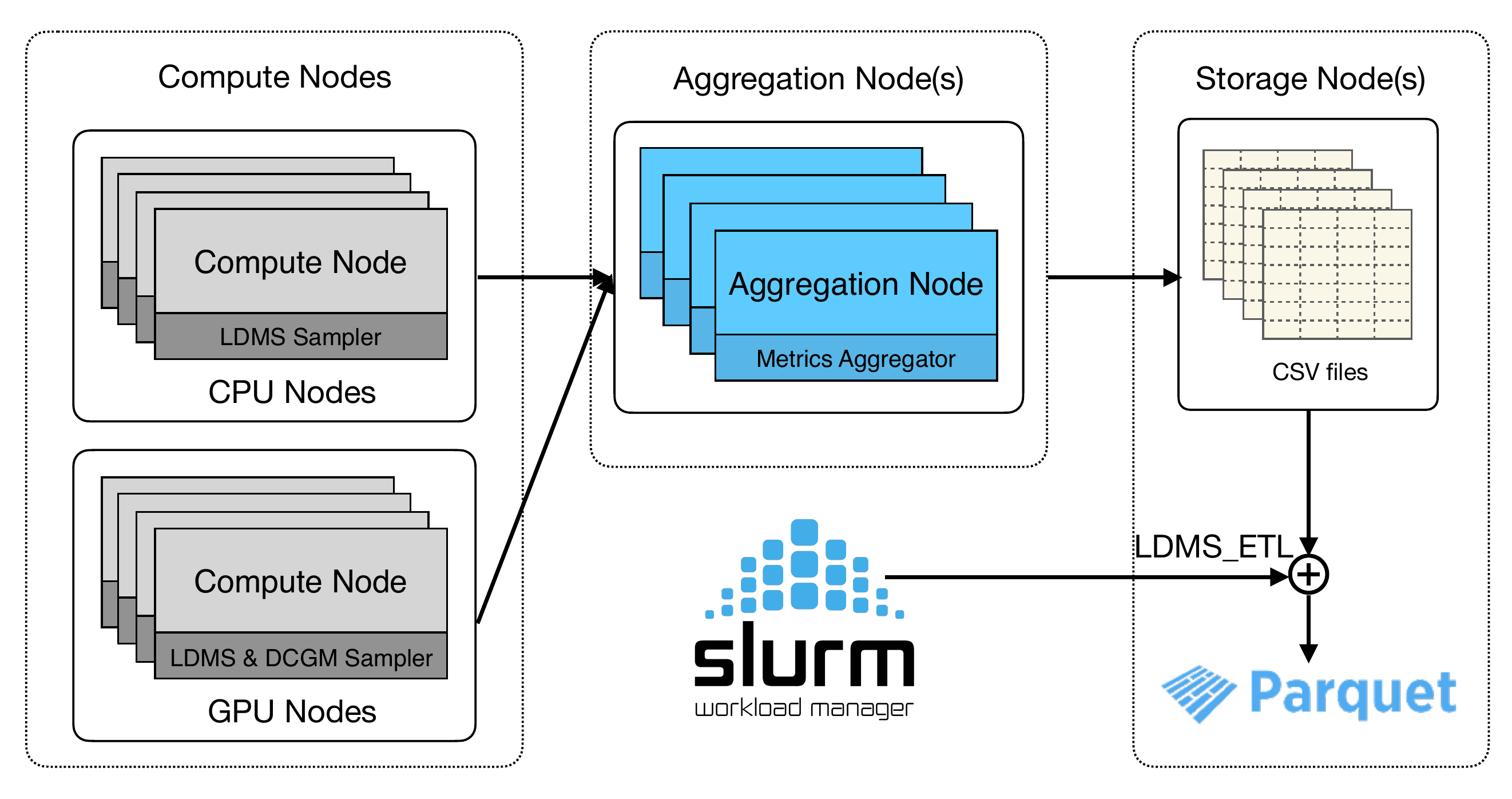}
    \caption{Data are collected from CPU-only and GPU nodes, aggregated by aggregation nodes, stored in CSV files, and then processed using python's parquet library after being joined by job-level data provided by SLURM.}
    \label{fig:data_collection}
\end{figure}

NERSC collects system-wide monitoring data through the Lightweight Distributed Metric Service (LDMS)~\cite{agelastos2014lightweight} and Nvidia's Data Center GPU Manager (DCGM)~\cite{dcgm}. LDMS is deployed on both CPU-only and GPU nodes; it samples node-level metrics either from a subset of hardware performance counters or operating system data, such as memory usage, I/O operations, etc. DCGM is dedicated to collecting GPU-specific metrics, including GPU utilization, GPU memory utilization, NVlink traffic, etc. The sampling interval of both LDMS and DCGM is set by the system at 10 seconds. The monitoring data are aggregated into CSV files from which we build a processing pipeline for our analysis, shown in Figure~\ref{fig:data_collection}. As a last step, we merge the job metadata from SLURM (job ID, job step, allocated nodes, start time, end time, etc.) with the node-level monitoring metrics. The output from our flow is a set of parquet files. 

Due to the large volume of data, we only sample Perlmutter from \emph{November 1} to \emph{December 1} of 2022. The system's monitoring infrastructure is still under deployment and some important traces such as memory bandwidth are not available at this time. A duration of one month is typically representative in an open-science HPC system~\cite{michelogiannakis2022case}, which we separately confirmed by sampling other periods. However, Perlmutter's workload may shift after the retirement of Cori as well as the introduction of policies such as allowing jobs to share nodes in a limited fashion. Still, a similar extensive study in Cori~\cite{michelogiannakis2022case} that allows node sharing extracted similar resource usage conclusions as our study. Therefore, we anticipate that the key insights from our study in Perlmutter will remain unchanged, and we consider that studies conducted in the early stages of a system's lifetime hold significant value.

We measure CPU utilization from \emph{cpu\_id} (CPU idle time among all cores in a node, expressed as a percentage) reported from \emph{vmstat} through LDMS~\cite{agelastos2014lightweight}; we then calculate CPU utilization (as a percentage) as: $100 - cpu\_id$. GPU utilization (as a percentage) is directly read from DCGM reports~\cite{dcgm_exporter}. Memory capacity utilization encompasses both the utilization of memory by user-space applications and the operating system. We use \emph{fb\_free} (framebuffer memory free) from DCGM to calculate GPU HBM2 utilization and \emph{mem\_free} (the amount of idle memory) from LDMS to calculate host DRAM capacity utilization. Memory capacity utilization (as a percentage) is calculated as $MemUtil = \frac{MemTotal - MemFree}{MemTotal} \times 100$, where $MemTotal$, as described above, is 512GB for CPU nodes, 256GB for the host memory of GPU nodes, and 40GB for each GPU HBM2. $MemFree$ is the unused memory of a node, which essentially shows how much more memory the job could have used.

In order to understand the temporal and spatial imbalance of resource usage among jobs, we use the equations proposed in~\cite{peng2021holistic} to calculate the temporal imbalance factor (\emph{$RI_{temporal}$}) and spatial imbalance factor (\emph{$RI_{spatial}$}). These factors allow us to quantify the imbalance in resource usage over time and across nodes, respectively. For a job that requests \emph{N} nodes and runs for time \emph{T}, and its utilization of resource \emph{$r$} on node \emph{n} at time \emph{t} is \emph{$U_{n,t}$}, the temporal imbalance factor is defined as:

\begin{equation}
     RI_{temporal}(r)  = \max_{1\leq n \leq N}(1 - \frac{\sum_{t=0}^{T} U_{n, t}}{\sum_{t=0}^{T}\max_{0\leq t \leq T}(U_{n, t})})
     \label{equ:temporal}
\end{equation}

\noindent Similarly, the spatial imbalance factor is defined as:

\begin{equation}
    RI_{spatial}(r) = 1- \frac{\sum_{n=1}^{N}\max_{0\leq t\leq T}(U_{n, t})}{\sum_{n=1}^{N}\max_{0\leq t\leq T, 1\leq n\leq N}(U_{n, t})}
    \label{equ:spatial}
\end{equation}

\noindent Both \emph{$RI_{temporal}$} and \emph{$RI_{spatial}$} are bound within the range of $[0, 1]$. Ideally, a job uses fully all resources on all allocated nodes across the job's lifetime, corresponding to a spatial and temporal factor of 0. A larger factor value indicates a variation in resource utilization temporally/spatially and the job experiences more temporal/spatial imbalance. 

\begin{table*}[!t]
    \centering
    \caption{Perlmutter measured data summary. Each job's resource utilization is represented by its peak usage.}
    \begin{tabularx}{\textwidth}{|c|*{4}{C}|*{4}{C}| }
        \hline
        \multirow{2}{*}{Metric} & \multicolumn{4}{c|}{\textbf{Statistics of all jobs }} & \multicolumn{4}{c|}{\textbf{Statistics of jobs $\geq 1h$}} \\
        \cline{2-9} 
        &  Median & Mean & Max & Std Dev & Median & Mean & Max & Std Dev\\
        \hline
        \rowcolor{Gray}
        & \multicolumn{4}{c}{\textbf{CPU Jobs}} & \multicolumn{4}{c|}{\textbf{21.75\% of CPU jobs} $\geq 1h$}\\
        \hline
        Allocated nodes & 1 & 6.51 & 1713 & 37.83 & 1 & 4.84 & 1477 & 25.43 \\
        \hline
        Job duration (hours) & 0.16 & 1.40 & 90.09 & 3.21 & 4.19 & 5.825 & 90.09 & 4.73 \\
        \hline
        CPU util (\%) & 35.0 & 39.98 & 100.0 & 34.60 & 51.0 & 56.68 & 100.0 & 35.89 \\
        \hline
        DRAM util (\%) & 13.29 & 22.79 & 98.62 & 23.65 & 18.61 & 33.69 & 98.62 & 30.88 \\
        \hline
        \rowcolor{Gray}
        & \multicolumn{4}{c}{\textbf{GPU Jobs}} & \multicolumn{4}{c|}{\textbf{23.42\% GPU jobs} $\geq 1h$}\\
        \hline
        Allocated nodes & 1 & 4.66 & 1024 & 27.71 & 1 & 5.88 & 512 & 23.33 \\
        \hline
        Job duration (hours) & 0.30 & 1.14 & 13.76 & 2.42 & 2.2 & 4.12 & 13.76 & 3.67 \\
        \hline
        Host CPU util (\%) & 4.0 & 19.60 & 100.0 & 23.53 & 4.0 & 18.00 & 100.0 & 24.81 \\
        \hline
        Host DRAM util (\%) & 17.57 & 29.76 & 98.29 & 12.51 & 18.04 & 28.24 & 98.29 & 20.94 \\
        \hline
        GPU util (\%) & 96.0 & 71.08 & 100.0 & 40.07 & 100.0 & 83.73 & 100.0 & 30.45 \\
        \hline
        GPU HBM2 util (\%) & 16.28 & 34.07 & 100.0 & 37.49 & 18.88 & 40.23 & 100.0 & 36.33 \\
        \hline
        \end{tabularx}
    \label{tab:data_summary}
\end{table*}

We exclude jobs with a runtime of less than 1 hour in our subsequent analysis, as such jobs are likely for testing or debugging purposes. Furthermore, since our sampling frequency is 10 seconds, it is difficult to capture peaks that last less than 10 seconds accurately. As a result, we concentrate on analyzing the behavior of sustained workloads. Table~\ref{tab:data_summary} summarizes job-level statistics in which each job's resource usage is represented by its maximum resource usage among all allocated nodes throughout its runtime.

\subsection{Analysis Methods}
To distill meaningful insights from our dataset we use \emph{Cumulative Distribution Functions (CDFs)}, \emph{Probability Density Functions (PDFs)}, and \emph{Pearson correlation coefficients}. The CDF shows the probability that the variable takes a value less than or equal to $x$, for all values of $x$; the PDF shows the probability that the variable has a value equal to $x$. To evaluate the resource utilization of jobs, we analyze the maximum resource usage that occurred during each job's entire runtime, and we factor in the job's impact on the system by weighting the job's data points based on the number of nodes allocated and the duration of the job. We then calculate the CDF and PDF of job-level metrics using these weighted data points. The Pearson correlation coefficient, which is a statistical tool to identify potential relationships between two variables, is used to investigate the correlation between two characteristics. The correlation factor, or Pearson's $r$, ranges from $-1.0$ to $1.0$; a positive value indicates a positive correlation, zero indicates no correlation, and a negative value indicates a negative correlation.

\section{Results}
\label{section:workload}

In this section, we start with an overview of the job characteristics, including their size, duration, and the applications they represent. Then we use CDF and PDF plots to investigate the resource usage pattern across jobs, followed by the characterization of the temporal and spatial variability of jobs. Lastly, we assess the correlation between the different resource types assigned to each job.

\subsection{Workloads Overview}
\label{subsection:overview}

\begin{table*}[!t]
    \centering
    \caption{Job size and duration. Jobs shorter than one hour are excluded.}
    \begin{tabularx}{\textwidth}{|c|c|*{6}{C}|}
        \hline
        \rowcolor{Gray}
        \multicolumn{2}{|c|}{\textbf{Job Size (Nodes)}} & \textbf{1} & \textbf{(1, 4]} & \textbf{(4, 16]} & \textbf{(16, 64]} & \textbf{(64, 128]} & \textbf{(128, 128+)} \\
        \hline
        \multirow{2}{*}{CPU Jobs} & Total Number: 21706 & 14783 & 2486 & 3738 & 550 & 62 & 87 \\ 
        \cline{2-8}
          & Percentage (\%) & \textbf{68.10} & 11.45	& 17.22 & 2.54 & 0.29 & 0.40 \\ 
        \hline
        \multirow{2}{*}{GPU Jobs} & Total Number: 24217 & 15924 & 5358 & 1837 & 706 & 318 & 74 \\
        \cline{2-8}
        & Percentage (\%) & \textbf{65.89} & 22.04 & 7.56 & 2.90 & 1.31 & 0.30 \\
        \hline
        
        \rowcolor{Gray}
        \multicolumn{2}{|c|}{\textbf{Job Duration (Hours)}} & \textbf{[1, 3]} & \textbf{(3, 6]} & \textbf{(6, 12]} & \textbf{(12, 24]} & \textbf{(24, 48]} & \textbf{(48, 48+)} \\
        \hline
        \multirow{2}{*}{CPU Jobs} & Total Number: 21706 & 8879 & 4109 & 6300 & 2393 & 15 & 10 \\ 
        \cline{2-8}
          & Percentage (\%) & \textbf{40.90} & 18.94	& 29.02 & 11.02 & 0.07 & 0.05 \\ 
        \hline
        \multirow{2}{*}{GPU Jobs} & Total Number: 24217 & 14495 & 3888 & 4916 & 918 & 0 & 0 \\
        \cline{2-8}
        & Percentage (\%) & \textbf{59.86} & 16.05 & 20.30 & 3.79 & 0 & 0 \\
        \hline
        
    \end{tabularx}
    \label{tab:job_size_distribution}
\end{table*}

We divide jobs into six groups by the number of allocated nodes and calculate the percentage of each group compared to the total number of jobs. The details are shown in Table~\ref{tab:job_size_distribution}. As shown, 68.10\% of CPU jobs and 65.89\% of GPU jobs only request one node, while large jobs that allocate more than 128 nodes are only 0.40\% and 0.30\% on CPU and GPU nodes, respectively. Also, 40.90\% of CPU jobs and 59.86\% of GPU jobs execute for less than three hours (as aforementioned, jobs with less than one hour of runtime are discarded from the dataset). We also observe that about 88.86\% of CPU jobs and 96.21\% of GPU jobs execute less than 12 hours, and only a few CPU jobs and no GPU jobs exceed 48 hours. This is largely a result of policy since Perlmutter's regular queue allows a maximum of 12 hours. However, jobs using a special reservation can exceed this limit~\cite{perlmutter}.

\begin{figure}[!t]
\centering
    \begin{subfigure}[b]{0.47\textwidth}
        \centering
        \includegraphics[width=\linewidth]{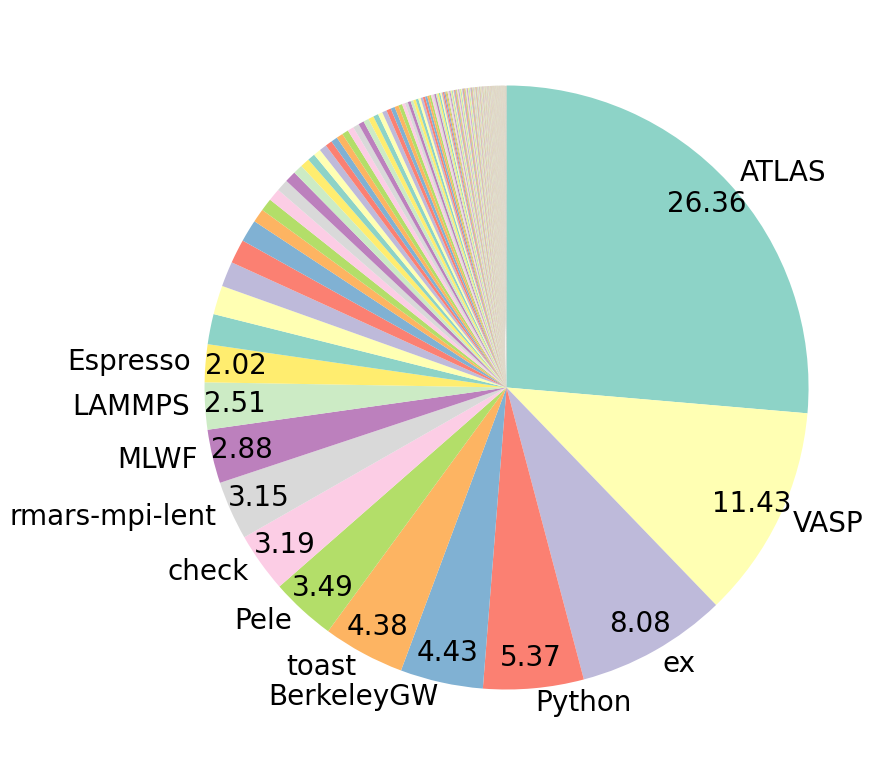}
        \caption{CPU-only nodes.} 
        \label{fig:cpu_app_distribution}
    \end{subfigure}%
    \begin{subfigure}[b]{0.5\textwidth}
        \centering
        \includegraphics[width=\linewidth]{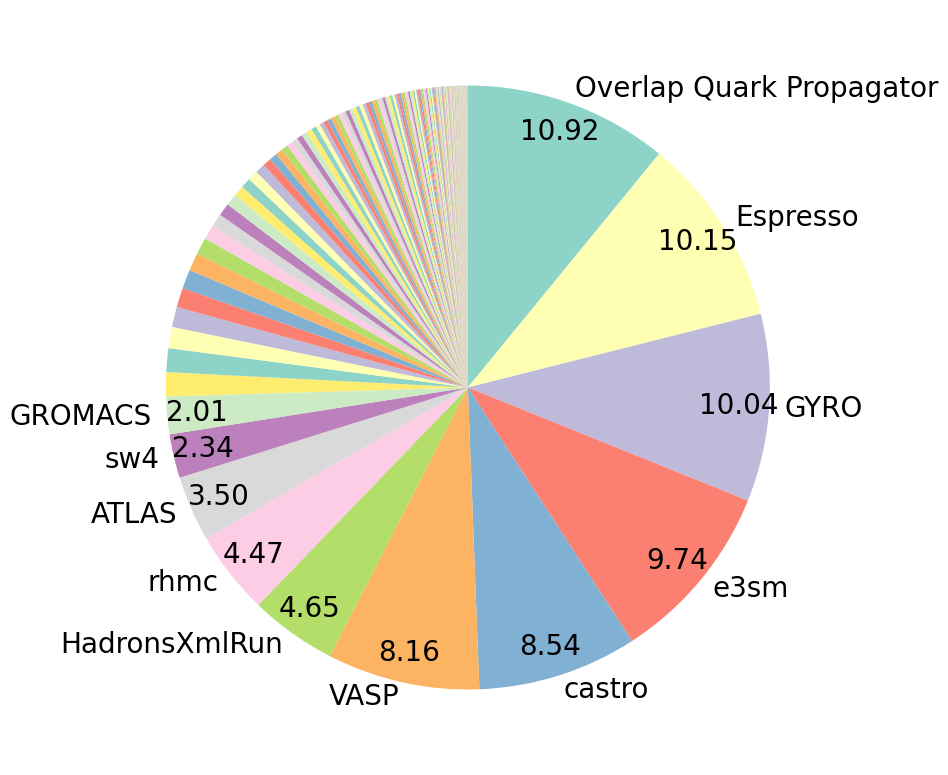}
        \caption{GPU-accelerated nodes.}
        \label{fig:gpu_app_distribution}
    \end{subfigure}
    \caption{Decomposition of node-hours by applications. Infrequent applications are not labeled.}
    \label{fig:apps}
\end{figure}

Next, we analyze the job names obtained from Slurm's \emph{sacct} and estimate the corresponding applications through empirical analysis. Although this approach has limitations, such as the inability to identify jobs with undescriptive names such as ``python'' or ``exec'', it still offers useful information. Figure~\ref{fig:apps} shows that most node hours on both CPU-only and GPU-accelerated nodes are consumed by a few recurring applications. The top four CPU-only applications account for 50\% of node hours, with ATLAS alone accounting for over a quarter. Over 600 CPU applications make up only 22\% of the node hours, using less than 2\% each (not labeled on the pie chart). On GPU-accelerated nodes, the top 11 applications consume 75\% of node hours, while the other 400+ applications make up the remaining 25\%. The top six GPU applications account for 58\% of node hours, with usage roughly evenly divided.

We further classify system workloads into three groups according to their maximum \emph{host} memory capacity utilization. In particular, jobs using less than 25\% of the total host memory capacity are categorized as low intensity, jobs that use 25-50\% are considered moderate intensity, and those exceeding 50\% are classified as high intensity~\cite{peng2021holistic}. Node-hours and the number of jobs can also be decomposed in these three categories, where node-hours is calculated by multiplying the total number of allocated nodes by the runtime (duration) of each job. 

\begin{figure}[!t]
\centering
    \begin{subfigure}[b]{0.5\textwidth}
        \centering
        \includegraphics[width=\linewidth]{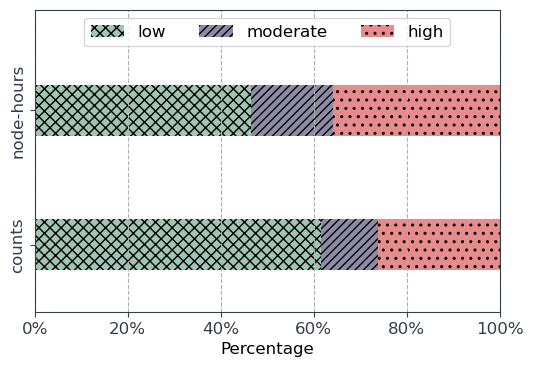}
        \caption{CPU-only jobs.} 
        \label{fig:cpu_job_decomposition}
    \end{subfigure}%
    \begin{subfigure}[b]{0.5\textwidth}
        \centering
        \includegraphics[width=\linewidth]{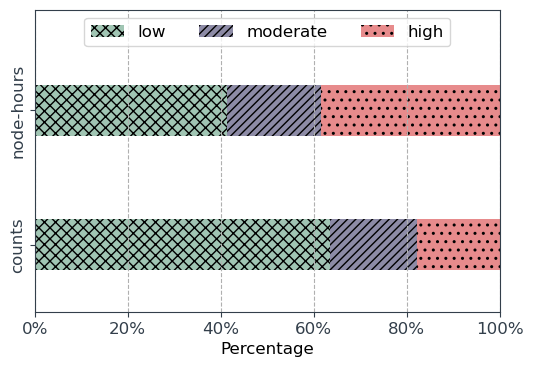}
        \caption{GPU-accelerated jobs.} 
        \label{fig:gpu_job_decomposition}
    \end{subfigure}
    \caption{Node-hours and job counts by host memory capacity intensity (utilization).}
\end{figure}

As shown in Figure~\ref{fig:cpu_job_decomposition}, CPU-only nodes have about 63\% of low memory capacity intensity jobs. Although moderate and high memory intensity jobs are 37\% of the total CPU jobs, they consume about 54\% of the total node-hours. This indicates that moderate and high memory intensity jobs are likely to use more nodes and/or run for a longer time. This observation holds true for GPU nodes in which 37\% of memory-intensive jobs compose 58\% of the total node-hours. In addition, we observe that even though the percentage of high memory intensity jobs on GPU nodes (17\%) is less than that on CPU nodes (26\%), the corresponding percentages of the node-hours are close, indicating that high memory intensity GPU jobs consume more nodes and/or run for a longer time than high memory intensity CPU jobs. 

\medskip
\noindent\fbox{%
    \parbox{0.97\linewidth}{%
        \textbf{Observation}: The analysis shows that both CPU and GPU nodes have around two-thirds of jobs that only occupy one node. GPU jobs have a higher proportion of short-lived jobs that run for less than three hours compared to CPU jobs. Additionally, jobs rarely allocate more than 128 nodes, which suggests that the majority of jobs can be accommodated within a single rack in the Perlmutter system. Furthermore, the analysis indicates that jobs that are intensive in host memory tend to consume more node-hours, despite representing a relatively small proportion of total jobs.
    }%
}

\subsection{Resource Utilization}
\label{subsection:resource}
This subsection analyzes resource usage among jobs and compares the characteristics of CPU-only jobs and GPU-enabled jobs. We consider the maximum resource usage of a job across all allocated nodes and throughout its entire runtime to represent its resource utilization because maximum utilization must be accounted for when scheduling a job in a system. As jobs with larger sizes and longer durations have a greater impact on system resource utilization, and the system architecture is optimized for node-hours, we calculate the resource utilization for each job and multiply the number of data points we add to our dataset that measure that utilization by the job's node-hours.

\subsubsection{CPU Utilization}

\begin{figure}[!t]
\centering
    \begin{subfigure}[b]{0.5\textwidth}
        \centering
        \includegraphics[width=\linewidth]{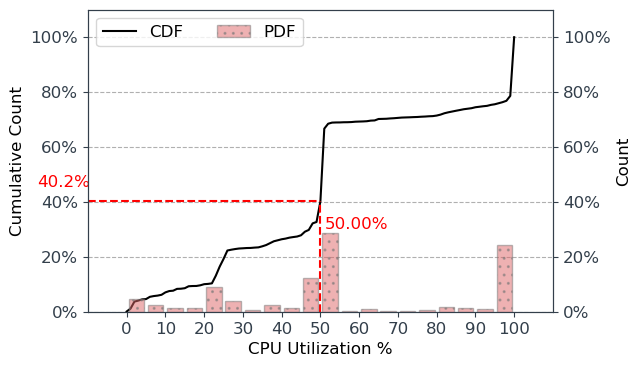}
    \end{subfigure}%
    \begin{subfigure}[b]{0.5\textwidth}
        \centering
        \includegraphics[width=\linewidth]{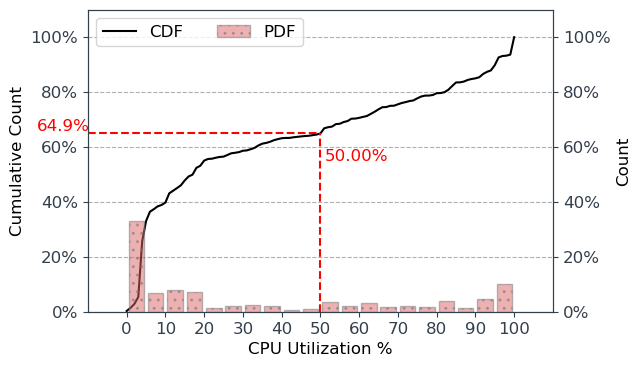}
    \end{subfigure}
    \caption{Maximum CPU utilization of CPU node-hours (left) and GPU node-hours (right).}
    \label{fig:job_cpu_util}
\end{figure}

Figure~\ref{fig:job_cpu_util} shows the distribution of the maximum CPU utilization of CPU jobs and GPU jobs weighted by node-hours. As shown, 40.2\% of CPU node-hours have at most 50\% CPU utilization, and about 28.7\% of CPU node-hours has a maximum CPU utilization of 50-55\%. In addition, 24.4\% of jobs reach over 95\% CPU utilization, creating a spike at the end of the CDF line. Over one-third of CPU jobs only utilize up to 50\% of the CPU resources available, which could potentially be attributed to Simultaneous Multi-threading (SMT) in the Milan architecture. While SMT can provide benefits for specific types of workloads, such as communication-bound or I/O-bound parallel applications, it may not necessarily improve performance for all applications and may even reduce it in some cases~\cite{tau2002empirical}. Consequently, users may choose to disable SMT, leading to half of the logical cores being unused during runtime. Additionally, certain applications are not designed to use SMT at all, resulting in a reported utilization of only 50\% in our analysis even with 100\% compute core utilization.

In contrast to CPU jobs, GPU-enabled jobs exhibit a distinct distribution of CPU usage, with the majority of jobs concentrated in the 0-5\% bin and only a small fraction of jobs utilizing the CPUs in full. We also obverse that node-hours with high utilization of both CPU and GPU resources are rare, with only 2.47\% of node-hours utilizing over 90\% of these resources (not depicted). This is because the CPUs in GPU nodes are primarily tasked with data preprocessing, data retrieval, and loading computed data, while the bulk of the computational load is offloaded to the GPUs. Therefore, the utilization of the CPUs in GPU-enabled jobs is comparatively low, as their primary function is to support and facilitate the GPU's heavy computational tasks.

\begin{figure}[!t]
\centering
    \begin{subfigure}[b]{0.5\textwidth}
        \centering
        \includegraphics[width=\linewidth]{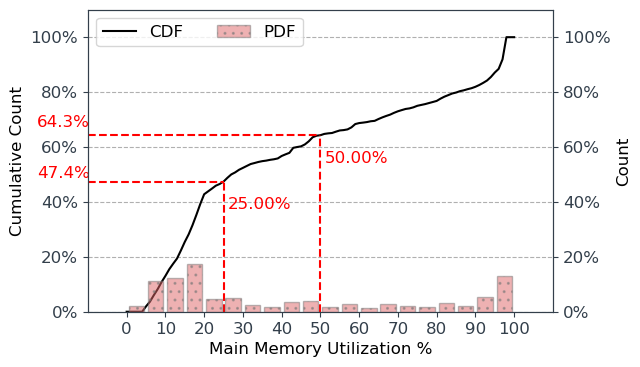}
    \end{subfigure}%
    \begin{subfigure}[b]{0.5\textwidth}
        \centering
        \includegraphics[width=\linewidth]{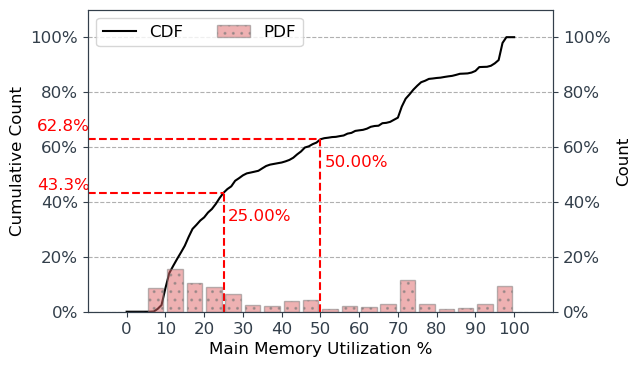}
    \end{subfigure}%
\caption{Maximum host memory capacity utilization of CPU node-hours (left) and GPU node-hours (right).}
\label{fig:job_mem_util}
\end{figure}

\subsubsection{Host DRAM Utilization} 
We plot the CDF and PDF of the maximum host memory utilization of job node-hours in Figure~\ref{fig:job_mem_util}. To help visualize the distribution of memory usage, the red vertical lines at the X axis indicate the 25\% and 50\% thresholds that we previously used to classify jobs into three memory intensity groups. A considerable fraction of the jobs on both CPU and GPU nodes use between 5\% and 25\% of host memory capacity, respectively. Specifically, 47.4\% of all CPU jobs and 43.3\% of all GPU jobs fall within these ranges. The distribution of memory utilization, like that of CPU utilization, displays spikes at the end of the CDF lines due to a small percentage of jobs (12.8\% for CPU and 9.5\% for GPU, respectively) that fully exhaust host memory capacity. 

Our results indicate that a significant proportion of both CPU and GPU jobs, 64.3\% and 62.8\% respectively, use less than 50\% of the available memory capacity. As a reminder, the available host memory capacity is 512 GB in CPU nodes and 256 GB in GPU nodes. While memory capacity is also not fully utilized in Cori~\cite{michelogiannakis2022case}, the higher memory capacity per node in Perlmutter exacerbates the challenge of fully utilizing the available memory capacity. 

\subsubsection{GPU Resources}
\begin{figure}[!t]
\centering
    \begin{subfigure}[b]{0.5\textwidth}
        \centering
        \includegraphics[width=\linewidth]{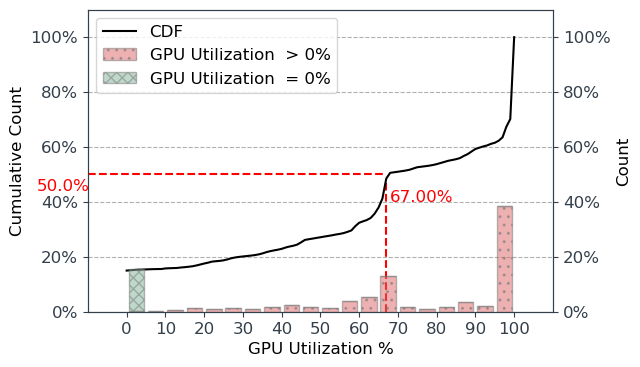}
    \end{subfigure}%
    \begin{subfigure}[b]{0.5\textwidth}
        \centering
        \includegraphics[width=\linewidth]{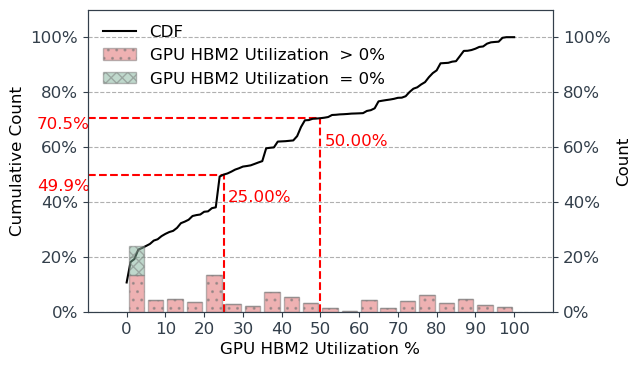}
    \end{subfigure}%
    \caption{Maximum GPU (left) and HBM2 capacity (right) utilization of GPU-hours.}
    \label{fig:gpu_gpu_hbm_util}
\end{figure}

The utilization of GPUs in DCGM indicates the percentage of time that GPU kernels are active during the sampling period, and it is reported per GPU instead of per node. Therefore, we analyze GPU utilization in terms of GPU-hours instead of node-hours. The left subfigure of Figure~\ref{fig:gpu_gpu_hbm_util} displays the CDF plot of maximum GPU utilization, indicating that 50\% of GPU jobs achieve a maximum GPU utilization of up to 67\%, while 38.45\% of GPU jobs reach a maximum GPU utilization of over 95\%. To assess the idle time of GPUs allocated to jobs, we separate the GPU utilization of zero from other ranges in the PDF histogram plot. As shown in the green bar, approximately 15\% of GPU hours are fully idle.

Similarly, we measure the maximum GPU HBM2 capacity utilization for each allocated GPU during the runtime of each job. As shown in the right subfigure of Figure~\ref{fig:gpu_gpu_hbm_util}, the HBM2 utilization is close to evenly distributed from 0\% to 100\%, resulting in a nearly linear CDF line. The green bar in the PDF plot suggests that 10.6\% of jobs use no HBM2 capacity, which is lower than the percentage of GPU idleness (15\%). This finding is intriguing as it indicates that even though some allocated GPUs are idle, their corresponding GPU memory is still utilized, possibly by other GPUs or for other purposes.

The GPU resources' idleness can be attributed to the current configuration of GPU-accelerated nodes, which are not allowed to be shared by jobs at the same time. As a result, each user has exclusive access to four GPUs per node, even if they require fewer resources. Sharing nodes may be enabled in the future, potentially leading to more efficient use of GPU resources.

\medskip
\noindent\fbox{%
    \parbox{0.97\linewidth}{%
        \textbf{Observation:} After analyzing CPU and host DRAM utilization, we find that GPU node-hours consume fewer CPU and host memory resources in comparison to CPU node-hours, likely because the computation is offloaded to GPUs. Although most GPU-hours reach high GPU utilization rates, we find that 15\% of them have fully idle GPUs, and 10.6\% of GPU-hours do not utilize HBM2 capacity, due to current configurations that do not allow for job sharing of GPU nodes. Allowing GPU sharing could alleviate the idleness of GPU resources and increase their average utilization.
    }%
}

\subsection{Temporal Characteristics}
\label{subsection:temporal}

Memory capacity utilization can become temporally imbalanced when a job does not utilize memory capacity evenly over time. Temporal imbalance is particularly common in applications that consist of phases that require different memory capacities. In such cases, a job may require significant amounts of memory capacity during some phases, while utilizing much less during others, resulting in a temporal imbalance of memory utilization.

We classify jobs into three patterns by the $RI_{temporal}$ value of host DRAM utilization: \emph{constant}, \emph{dynamic}, and \emph{sporadic}~\cite{peng2021holistic}. Jobs with $RI_{temporal}$ lower than 0.2 are classified in the \emph{constant} pattern, where memory utilization does not show significant change over time. Jobs with $RI_{temporal}$ between 0.2 and 0.6 are in the \emph{dynamic} pattern, where jobs have frequent and considerable memory utilization changes. The \emph{sporadic} pattern is defined by $RI_{temporal}$ larger than 0.6. In this pattern, jobs have infrequent and sporadic higher memory capacity usage than the rest of the time.

\begin{figure}[!t]
\centering
    \begin{subfigure}[b]{0.33\textwidth}
        \centering
        \includegraphics[width=\linewidth]{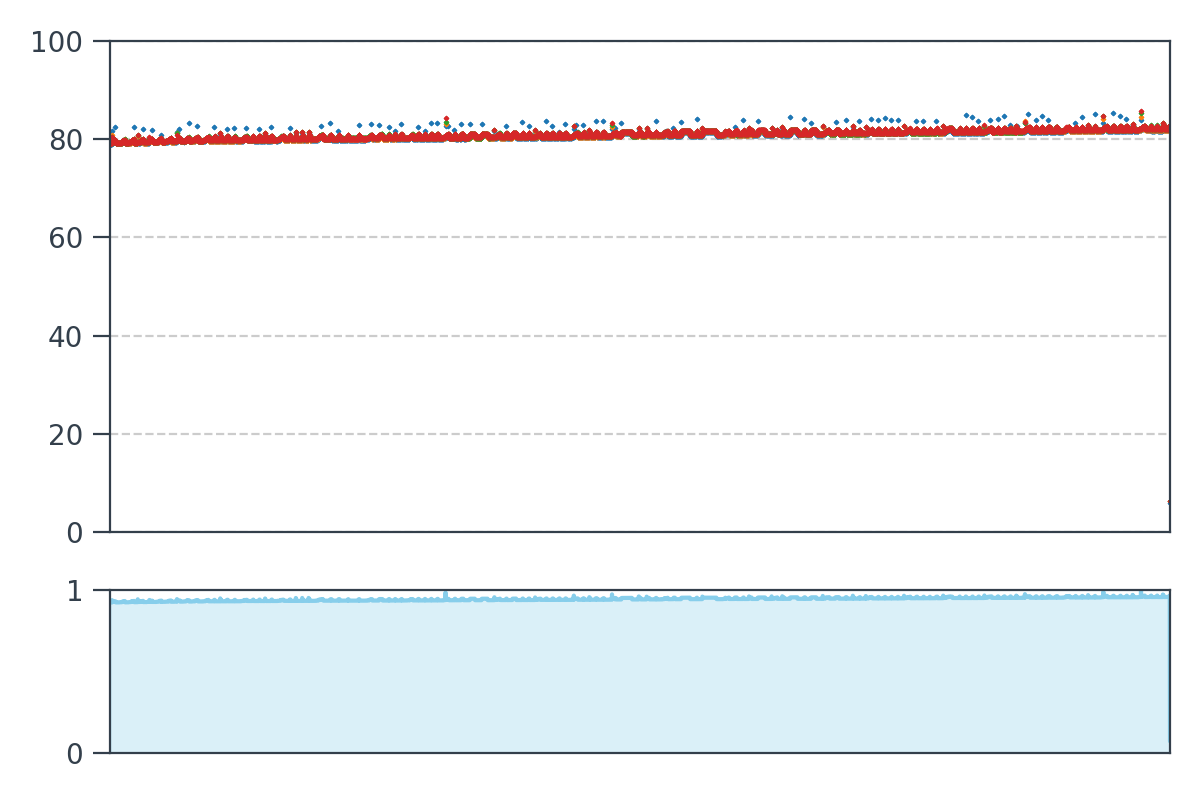}
        \caption{Constant pattern.}
    \end{subfigure}
    \begin{subfigure}[b]{0.33\textwidth}
        \centering
        \includegraphics[width=\linewidth]{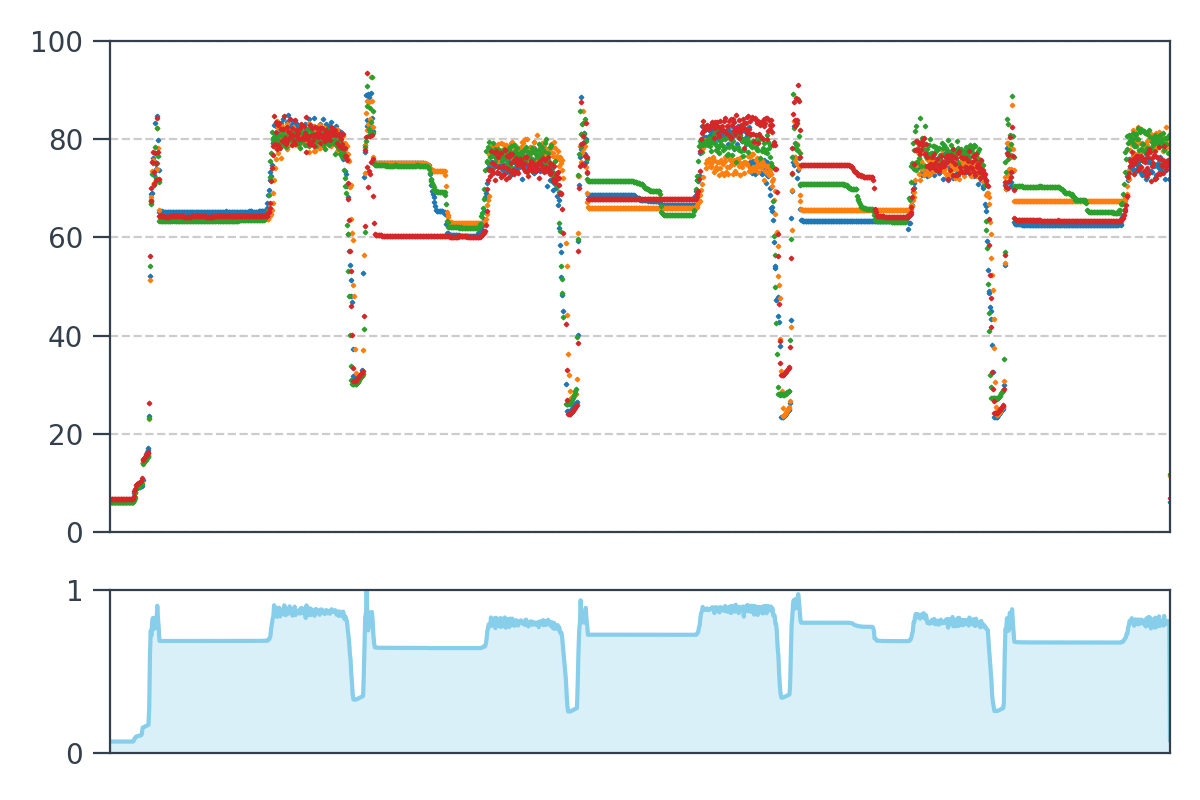}
        \caption{Dynamic pattern.}
        \end{subfigure}
    \begin{subfigure}[b]{0.33\textwidth}
        \centering
        \includegraphics[width=\linewidth]{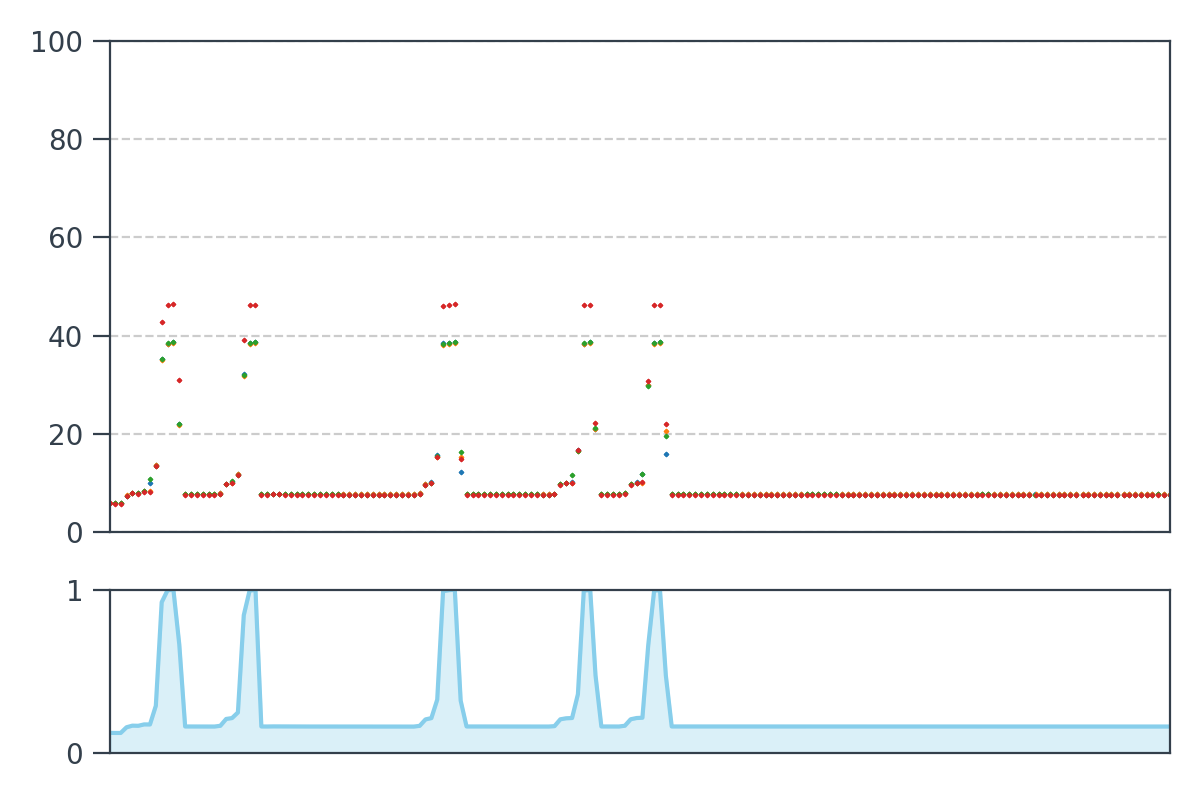}
        \caption{Sporadic pattern.}
    \end{subfigure}
\caption{Temporal patterns illustrated with the memory capacity utilization metrics of randomly selected jobs in Perlmutter, one representative job for each of the three categories. Each color represents the memory capacity utilization (\%) of each node assigned to the job over the job's runtime. The area plots at the bottom show the normalized metrics for the node that has the maximum temporal factor among nodes allocated to the job; the percentage of the blank area corresponds to the value of $RI_{temporal}$ of a job. A larger blank area indicates more temporal imbalance.}
\label{fig:job_mem_util_ts_example}
\end{figure}

Figure~\ref{fig:job_mem_util_ts_example} illustrates three memory utilization patterns that were constructed from our monitoring data. Each color in the scatter plot represents a different node allocated to the job. The constant pattern job shows a nearly constant memory capacity utilization of about 80\% across all allocated nodes for its entire runtime, resulting in the bottom area plot being almost fully covered. The dynamic pattern job also exhibits similar behavior across its allocated nodes, but due to variations over time, the shaded area has several bumps and dips, resulting in an increase in the blank area. For the sporadic pattern job, the memory utilization readings of all nodes have the same temporal pattern, with sporadic spikes and low memory capacity usage between spikes, resulting in the blank area occupying most of the area and indicating poor temporal balance.

\begin{figure}[!t]
\centering
    \begin{subfigure}[b]{0.5\textwidth}
        \centering
        \centering
        \includegraphics[width=\linewidth]{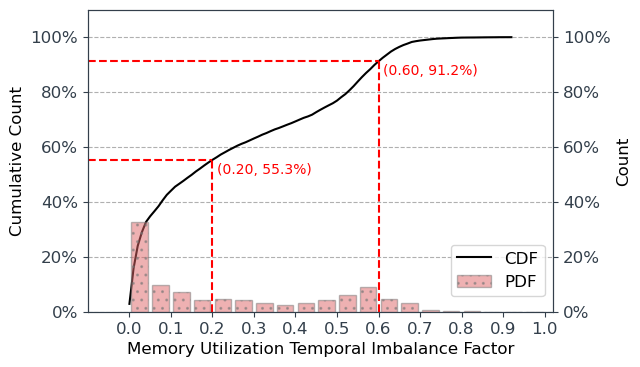}
        \caption{CPU jobs.}
        \label{fig:cpu_job_mem_util_temporal}
    \end{subfigure}%
    \begin{subfigure}[b]{0.5\textwidth}
        \centering
        \includegraphics[width=\linewidth]{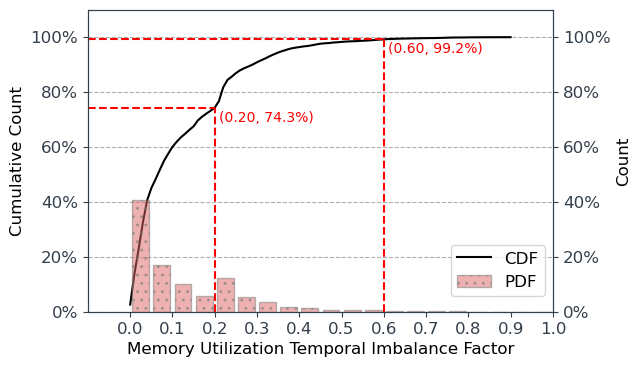}
        \caption{GPU jobs.}
        \label{fig:gpu_job_mem_util_temporal}
    \end{subfigure}
    \caption{CDFs and PDFs of the \emph{temporal} factor of host memory capacity utilization across nodes. The larger the value of the temporal factor, the more temporal imbalance.}
     \label{fig:job_mem_util_temporal}
\end{figure}

The CDFs and PDFs of the host memory temporal imbalance factor of CPU jobs and GPU jobs are illustrated in Figure~\ref{fig:job_mem_util_temporal}, in which two vertical red lines separate the jobs into three temporal patterns. Overall, both CPU jobs and GPU jobs have good temporal balance: 55.3\% of CPU jobs and 74.3\% of GPU jobs belong to the constant pattern, i.e, their $RI_{temporal}$ values are below 0.2. Jobs on CPU nodes have a higher percentage of dynamic patterns: 35.9\% of CPU jobs have $RI_{temporal}$ value between 0.2 and 0.4, while GPU jobs have 24.9\% in the dynamic pattern. On GPU nodes, we only observe very few jobs (0.8\%) in the sporadic pattern, which means the cases of host DRAM having severe temporal imbalance are few.

\begin{figure}[!t]
\centering
    \begin{subfigure}[b]{0.5\textwidth}
        \centering
        \includegraphics[width=0.5\linewidth]{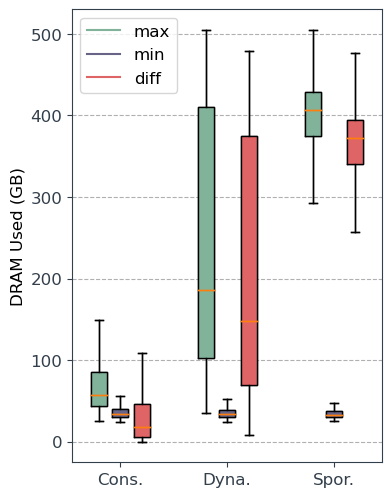}%
        \hfill
        \includegraphics[width=0.5\linewidth]{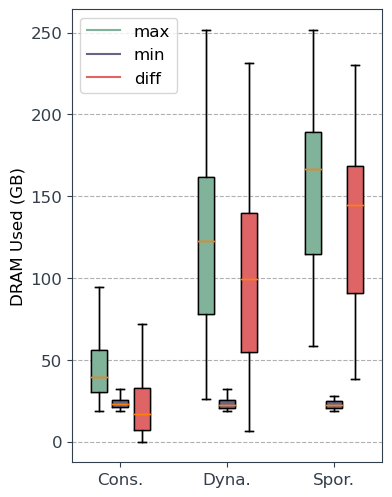}
        \caption{Temporal categories.} 
        \label{fig:job_mem_temporal}
    \end{subfigure}%
    \begin{subfigure}[b]{0.5\textwidth}
        \centering
        \includegraphics[width=0.5\linewidth]{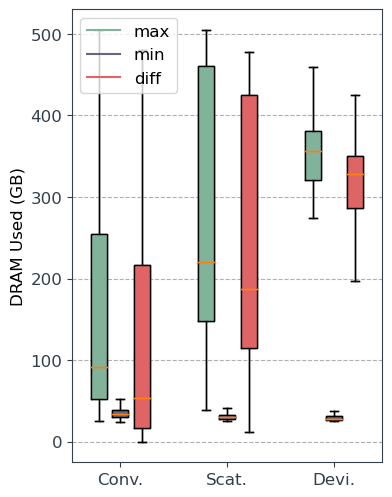}%
        \hfill
        \includegraphics[width=0.5\linewidth]{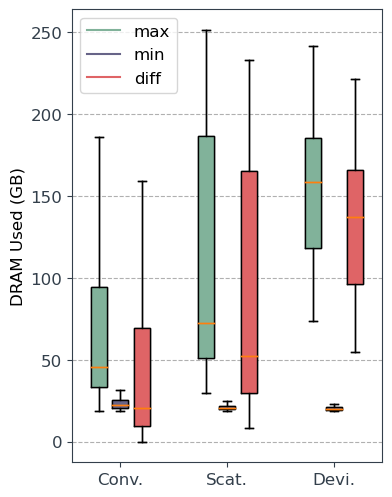}
        \caption{Spatial categories.} 
        \label{fig:job_mem_spatial}
    \end{subfigure}%
\caption{Host DRAM distribution by temporal and spatial categories. The left portion of each subfigure represents CPU jobs and the right portion GPU jobs.}
\end{figure}

We further analyze the memory capacity utilization distribution of jobs in each temporal pattern; the results are shown in Figure~\ref{fig:job_mem_temporal}. We extract the maximum, minimum, and difference between maximum and minimum memory capacity used from jobs in each category and present the distribution in box plots. The minimum memory used for all categories on the same nodes is similar: about 25 GB and 19 GB on CPU and GPU nodes, respectively. 75\% of jobs in the constant category on CPU nodes use less than 86 GB while 75\% jobs on GPU nodes use less than 56 GB. As 55.3\% CPU jobs and 74.3\% GPU jobs are in the constant category, 41.5\% CPU jobs and 55.7\% GPU jobs do not use 426 GB and 200 GB of the available capacity, respectively. The maximum memory used in the constant pattern is 150 GB on CPU nodes and 94 GB on GPU nodes, both of which do not exceed half of the memory capacity. Jobs using high memory capacity are only observed in dynamic and sporadic patterns, where 75\% sporadic jobs use up to 429 GB on CPU nodes and 189 GB on GPU nodes, respectively.

\medskip
\noindent\fbox{%
    \parbox{0.97\linewidth}{%
        \textbf{Observation}: Our analysis suggests that GPU nodes exhibit a greater proportion of jobs with temporal balance in host DRAM usage compared to CPU nodes. While over half of both CPU and GPU jobs fall under the category of temporal constant jobs, jobs with temporal imbalance, characterized by dynamic and sporadic patterns, generally require higher maximum memory capacity compared to constant pattern jobs. Furthermore, the distribution of host memory capacity usage among jobs with different temporal patterns reveals that memory capacity is not fully utilized for constant pattern jobs, whereas dynamic and sporadic pattern jobs may achieve high memory capacity utilization at some point during their runtime.
    }%
}

\subsection{Spatial Characteristics}
\label{subsection:spatial}

\begin{figure}[!t]

    \begin{subfigure}[b]{0.33\textwidth}
    \includegraphics[width=\linewidth]{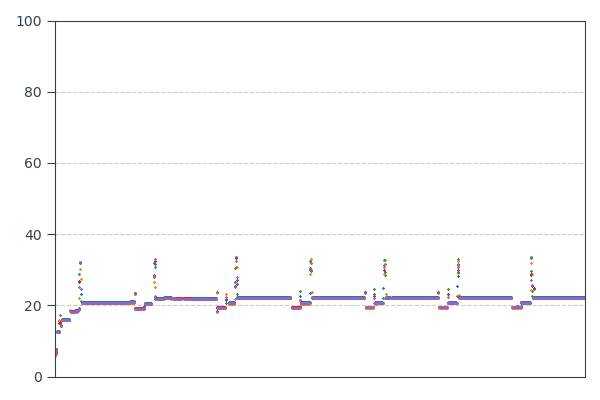}
    \caption{Convergent pattern.}
    \end{subfigure}%
    \begin{subfigure}[b]{0.33\textwidth}
    \includegraphics[width=\linewidth]{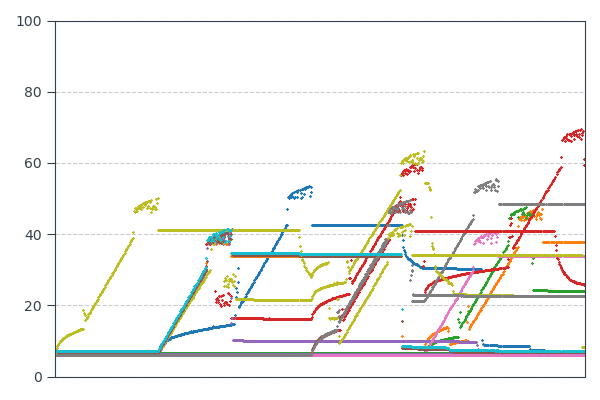}
    \caption{Scattered pattern.}
    \end{subfigure}%
    \begin{subfigure}[b]{0.33\textwidth}
    \includegraphics[width=\linewidth]{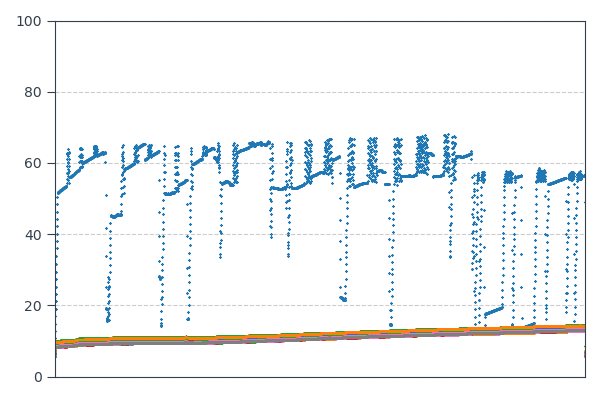}
    \caption{Deviational pattern.}
    \end{subfigure}
    \caption{Spatial patterns illustrated with the memory capacity utilization metrics of randomly selected jobs in Perlmutter, one representative job for each of the three categories. Each color represents memory utilization (\%) of a different node allocated to each job.}
    \label{fig:job_mem_util_sp_example}
\end{figure}

The job scheduler and resource manager of current HPC systems do not consider the varying resource requirements of individual tasks within a job, leading to spatial imbalances in resource utilization across nodes. One common type of spatial imbalance is when a job requires a significant amount of memory in a small number of nodes, while other nodes use relatively less memory. Spatial imbalance of memory capacity quantifies the uneven usage of memory capacity across nodes allocated to a job. 

To characterize the spatial imbalance of jobs, we use equation~\ref{equ:spatial} presented in~\ref{section:data collection} to calculate the spatial factor $RI\_spatial$ of memory capacity usage for each job. Similar to the temporal factor, $RI\_spatial$ falls in the range [0, 1] and larger values represent higher spatial imbalance. Jobs are classified into one of three spatial patterns: (i) \emph{convergent} pattern that has $RI\_spatial$ less than 0.2, (ii) \emph{scattered} pattern that has $RI\_spatial$ between 0.2 and 0.6, and (iii) \emph{deviational} pattern with its $RI\_spatial$ larger than 0.6. 

\begin{figure}[!t]
\centering
    \begin{subfigure}[b]{0.5\textwidth}
        \centering
        \includegraphics[width=\linewidth]{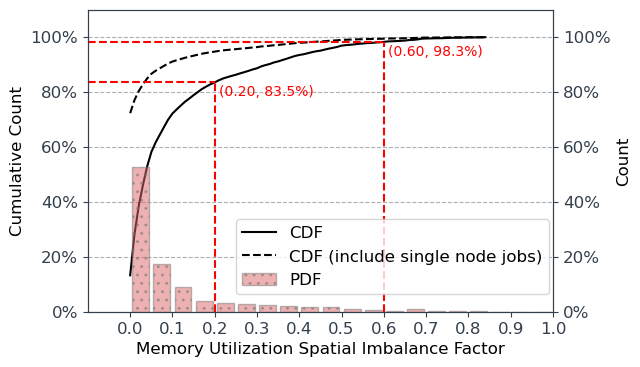}
        \caption{CPU jobs.}
        \label{fig:cpu_job_mem_util_spatial}
    \end{subfigure}%
    \begin{subfigure}[b]{0.5\textwidth}
        \centering
        \includegraphics[width=\linewidth]{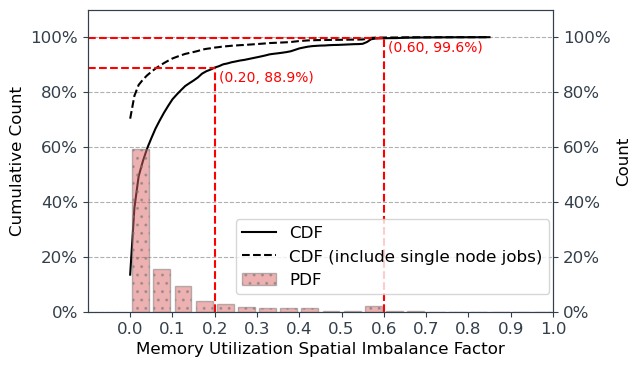}
        \caption{GPU jobs.}
        \label{fig:gpu_job_mem_util_spatial}
    \end{subfigure}%
\caption{CDFs and PDFs of the \emph{spatial} factor of host memory capacity utilization of jobs. The larger the value of the spatial factor, the more spatial imbalance.} 
\label{fig:job_mem_util_spatial}
\end{figure}

As shown in the examples in Figure~\ref{fig:job_mem_util_sp_example}, a job that exhibits a convergent pattern has similar or identical memory capacity usage among all of its assigned nodes. A job with a scattered pattern shows diverse memory usage and different peak memory usage among its nodes. A spatial deviational pattern job has a similar memory usage pattern in most of its nodes but has one or several nodes deviate from the bunch. It is worth noting that low spatial imbalance does not indicate low temporal imbalance. The spatial convergent pattern job shown in the example has several spikes in memory usage and therefore is a temporal sporadic pattern.

We present the CDFs and PDFs of the job-wise host memory capacity spatial factor in Figure~\ref{fig:job_mem_util_spatial}. Overall, 83.5\% of CPU jobs and 88.9\% of GPU nodes are in the convergent pattern and very few jobs are in the deviational pattern. Because jobs that allocate a single node always have a spatial imbalance factor of zero, if we include single-node jobs, the overall memory spatial balance is even better: 94.7\% for CPU jobs and 96.2\% for GPU jobs. 

We combine the host memory spatial pattern with the host memory capacity usage behavior in each job and plot the distribution of memory capacity utilization by spatial patterns; the results are shown in Figure~\ref{fig:job_mem_spatial}. Similar to the distribution of the temporal patterns, we use the maximum, minimum, and difference of job memory to evaluate the memory utilization imbalance. Spatial convergent jobs have relatively low memory usage. As shown in the green box plots, 75\% of spatial convergent jobs (upper quartile) use less than 254 GB on CPU nodes and 95 GB on GPU nodes. Given that spatial convergent jobs account for over 94\% of total jobs, over 70\% of jobs have 258 GB and 161 GB of memory capacity unused for CPU and GPU nodes, respectively. Memory imbalance, i.e, the difference between the maximum and minimum memory capacity usage of a job (red box plots), is also the lowest in convergent pattern jobs. For spatial-scattered jobs on CPU nodes, even though they are a small portion of the total jobs, the memory difference spans a large range: from 115 GB at 25\% percentile to 426 GB at 75\% percentile. Spatial deviational CPU jobs have a shorter span in memory imbalance compared to GPU jobs; it only ranges from 286 GB to 350 GB at the lower and upper quartiles, respectively.

\medskip
\noindent\fbox{%
    \parbox{0.97\linewidth}{%
        \textbf{Observation}: Our analysis shows that a significant number of CPU and GPU jobs on Perlmutter have a convergent pattern of spatial balance for host memory capacity usage across allocated nodes. Even after eliminating single-node jobs, the proportion of jobs with a convergent spatial pattern remains high, suggesting that Perlmutter's jobs generally have good spatial balance. However, jobs with scattered and deviational spatial patterns, albeit fewer in number, tend to consume more memory capacity in some allocated nodes, leading to uneven memory capacity utilization across nodes and some nodes exhibiting low memory capacity utilization.
    }%
}

\subsection{Correlations}

\begin{figure}[!t]
\centering
    \begin{subfigure}[b]{0.5\textwidth}
    \centering
    \includegraphics[width=\linewidth]{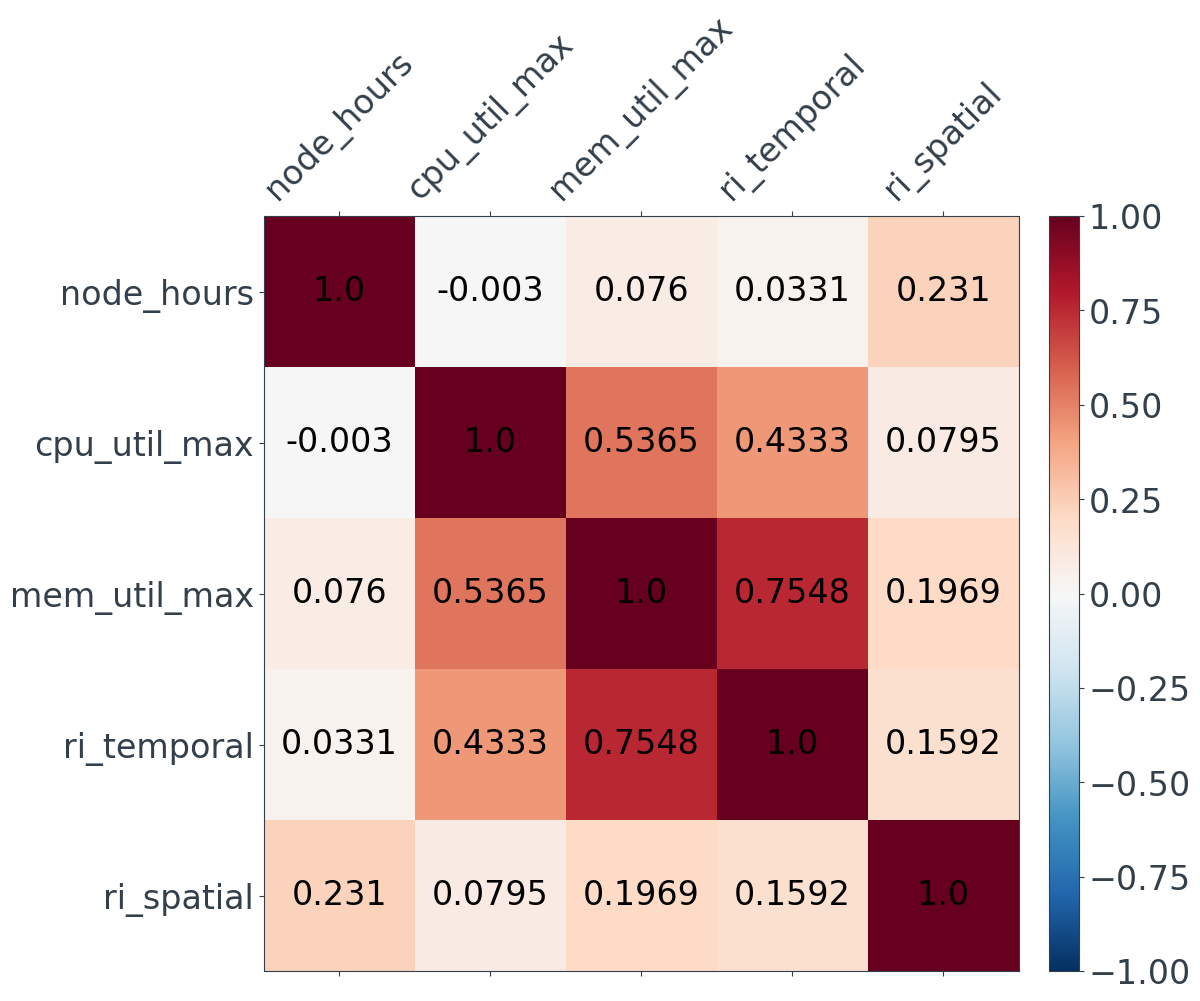}
    \caption{CPU jobs.}
    \end{subfigure}%
    \begin{subfigure}[b]{0.5\textwidth}
    \centering
    \includegraphics[width=\linewidth]{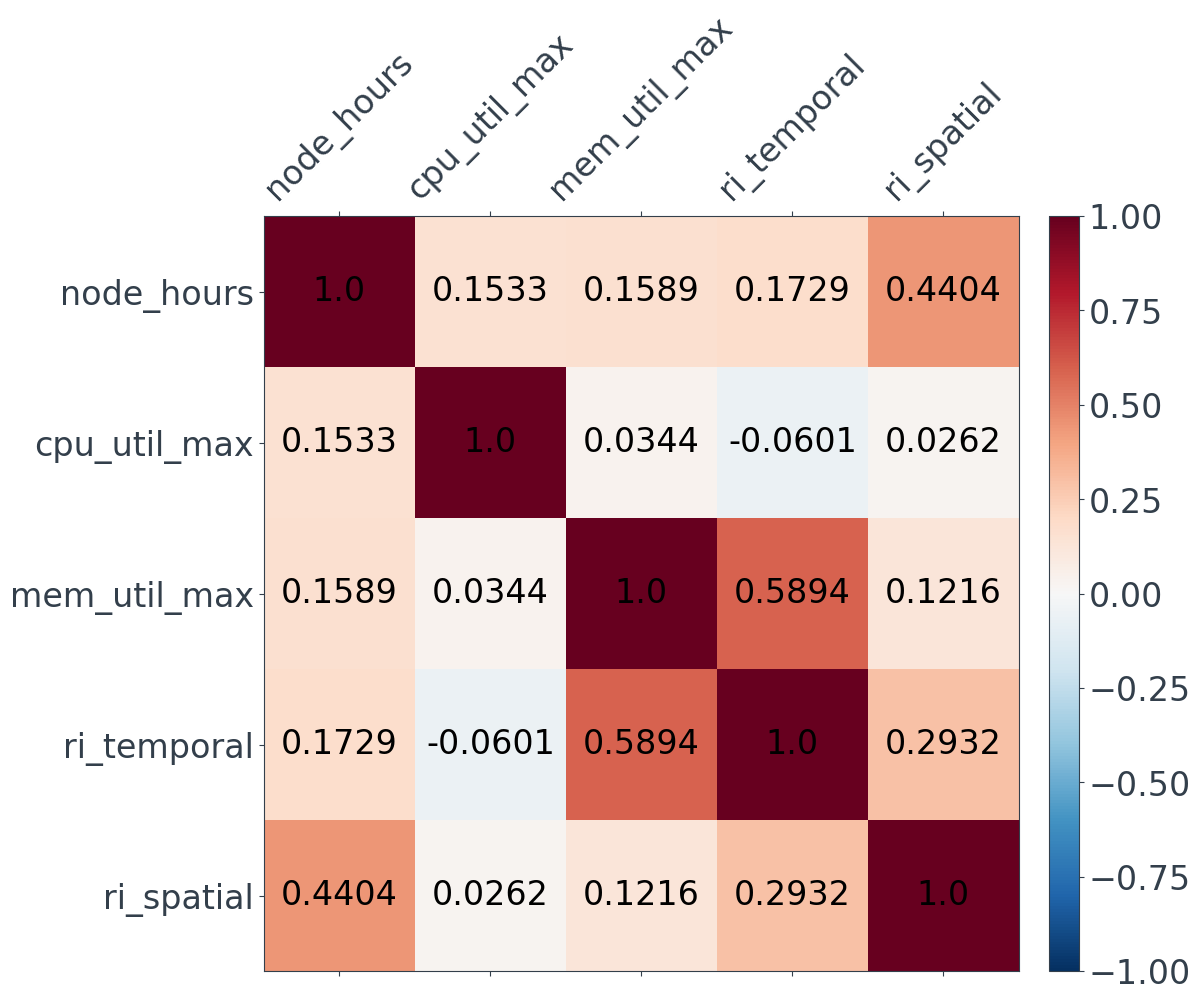}
    \caption{GPU jobs.}
    \end{subfigure}
    \caption{Correlation of job node-hours, maximum memory capacity used, temporal, and spatial factors.}
    \label{fig:job_correlation}
\end{figure}

We conduct an analysis of the relationships between various job characteristics on Perlmutter, including job size and duration (measured as $node\_hours$), maximum CPU and host memory capacity utilization, and temporal and spatial factors. The results of the analysis are presented in a correlation matrix in Figure~\ref{fig:job_correlation}. Our findings show that for both CPU and GPU nodes, job node-hours are positively correlated with the spatial imbalance factor ($ri\_spatial$). This suggests that larger jobs with longer runtimes are more likely to experience spatial imbalance. Maximum CPU utilization is strongly positively correlated with host memory capacity utilization and temporal factors in CPU jobs, while the correlation is weak in GPU jobs. Moreover, the temporal imbalance factor ($ri\_temporal$) is positively correlated with maximum memory capacity utilization ($mem\_max$), with correlation coefficients (r-value) of 0.75 for CPU jobs and 0.59 for GPU jobs. These strong positive correlations suggest that jobs requiring a significant amount of memory are more likely to experience temporal memory imbalance, which is consistent with our previous observations. Finally, we find a slight positive correlation (r-value of 0.16 for CPU jobs and 0.29 for GPU jobs) between spatial and temporal imbalance factors, indicating that spatially imbalanced jobs are also more likely to experience temporal imbalance.

\section{Discussion and Conclusion}
\label{section:conclusion}

In light of the increasing demands of HPC and the varied resource requirements of open-science workloads, there is a risk of not fully utilizing expensive resources. To better understand this issue, we conducted a comprehensive analysis of memory, CPU, and GPU utilization in NERSC's Perlmutter. Our analysis spanned \emph{one month} and yielded important insights. Specifically, we found that only a quarter of CPU node-hours achieved high CPU utilization, and CPUs on GPU-accelerated nodes were typically utilized for only 0-5\% of the node-hours. Moreover, while a significant proportion of GPU-hours demonstrated high GPU utilization (over 95\%), more than 15\% of GPU-hours had idle GPUs. Moreover, both CPU host memory and GPU HBM2 were not fully utilized for the majority of node-hours. Interestingly, jobs with temporal balance consistently did not fully utilize memory capacity, while those with temporal imbalance had varying idle memory capacity over time. Finally, we observed that jobs with spatial imbalance did not have high memory capacity utilization for all allocated nodes.

Insufficient resource utilization can be attributed to various application characteristics, as similar issues have been observed in other HPC systems. Although simultaneous multi-threading can potentially improve CPU utilization and mitigate stalls resulting from cache misses, it may not be suitable for all applications. Furthermore, GPUs, being a new compute resource to NERSC users, may be currently not fully utilized because users and applications are still adapting to the new system, and the current configurations are not optimized yet to support GPU node sharing. Furthermore, it is important to note that in most systems, various parameters such as memory bandwidth and capacity are interdependent. For instance, the number and type of memory modules significantly impact memory bandwidth and capacity. Therefore, when designing a system, it may be challenging to fully utilize every parameter while optimizing others. This may result in some resources being not fully utilized to improve the overall performance of the system. Thus, not fully utilizing system resources can be an intentional trade-off in the design of HPC systems. 

Our study provides valuable insights for system operators to understand and monitor resource utilization patterns in HPC workloads. However, the scope of our analysis was limited by the availability of monitoring data, which did not include information on network and memory bandwidth as well as file system statistics. Despite this limitation, our findings can help system operators identify areas where resources are not fully utilized and optimize system configuration.

Our analysis also reveals several opportunities for future research. For instance, given that 64\% of jobs use only half or less of the on-node host DRAM capacity, it is worth exploring the possibility of disaggregating the host memory and using a remote memory pool. This remote pool can be local to a rack, group of racks, or the entire system. Our job size analysis indicates that most jobs can be accommodated within the compute resources provided by a single rack, suggesting that rack-level disaggregation can fulfill the requirements of most Perlmutter jobs if they are placed in a single rack. Furthermore, a disaggregated system could consider temporal and spatial characteristics when scheduling jobs since high memory utilization is often observed in memory-unbalanced jobs. Such jobs can be given priority for using disaggregated memory.

Another promising area for improving resource utilization is to reevaluate node sharing for specific applications with compatible temporal and spatial characteristics. One of the main challenges in job co-allocation is the potential for shared resources, such as memory, to become saturated at high core counts and significantly degrade job performance. However, our analysis reveals that both CPU and memory resources are not fully utilized, indicating that there may be room for co-allocation without negatively impacting performance. The observation that memory-balanced jobs typically consume relatively low memory capacity suggests that it may be possible to co-locate jobs with memory-balanced jobs to reduce the probability of contention for memory capacity. By optimizing resource allocation and reducing the likelihood of resource contention, these approaches can help maximize system efficiency and performance.

\section*{Acknowledgment}


We would like to express our gratitude to the anonymous reviewers for their insightful comments and suggestions. We also thank Brian Austin, Nick Wright, Richard Gerber, Katie Antypas, and the rest of the NERSC team for their feedback. This research used resources of the National Energy Research Scientific Computing Center (NERSC), a U.S. Department of Energy Office of Science User Facility located at Lawrence Berkeley National Laboratory, operated under Contract No. DE-AC02-05CH11231. This work was supported by the Director, Office of Science, of the U.S. Department of Energy under Contract No. DE-AC02-05CH11231. 
This research was supported in part by the National Science Foundation under grants OAC-1835892 and CNS-1817094.


%
%
\bibliographystyle{splncs04}
\bibliography{bibliography}
\end{document}